\documentclass[12pt,a4paper]{cibb}

\makeatletter
\providecommand{\@ordinalM}[2]{#1}
\makeatother

\usepackage{subfigure,graphicx}
\usepackage{amsmath,amsfonts,latexsym,amssymb,euscript,xr}
\usepackage{booktabs}
\usepackage[nodayofweek]{datetime}
\usepackage{hyperref}
\usepackage{fmtcount}
\usepackage[english]{datenumber}
\usepackage[absolute]{textpos}

\usepackage[table]{xcolor}
\usepackage{color,colortbl,tabularx}

\usepackage[english]{babel}
\usepackage[protrusion=true,expansion=true]{microtype}
\usepackage{amsmath,amsfonts,amsthm}
\usepackage{pifont}

\definecolor{LightBlue}{rgb}{0.88,0.9,0.9}

\newcommand{\party}{\mathcal{P}} 
\newcommand{\helper}{Helper} 

\title{\Large $\ $\\ \bf Privacy-Preserving Detection of Rare Disease-Associated Cell Subsets via Secure Multi-Party Computation}

\author{\large Ş. Selcan Magara$^{*,1,2}$, Esther Havemann$^{*1,2}$, Debora Jutz$^{3}$, Ali Burak Ünal$^{4}$ and Mete Akgün$^{1,2}$}
\address{ \footnotesize $\ $\\$^1$ Institute for Bioinformatics and Medical Informatics, University of Tübingen, Tübingen, Germany \\
$^2$ Medical Data Privacy and Privacy-Preserving Machine Learning, University of Tübingen, Tübingen, Germany \\
$^3$ Lab for Artificial Intelligence in Medicine, Department of Diagnostic and Interventional Radiology, University Hospital Aachen, Aachen, Germany  \\
$^4$ Intelligent Vehicles Lab, Delft University of Technology, Delft, Netherlands \\
\bigskip
ORCID code(s): ŞSM 0000-0002-0427-811X.
\bigskip
\newline
$^*$corresponding authors: seyma-selcan.magara@uni-tuebingen.de, esther.havemann@student.uni-tuebingen.de
\bigskip
\newline
}

\abstract{\small single-cell analysis, multi-party computation, privacy-preserving machine learning \normalsize
\\[17pt]
{\bf Abstract} The detection of rare disease-associated cell subsets from high-dimensional single-cell measurements is critical for understanding diseases such as leukaemia and viral infections. CellCnn, a convolutional neural network (CNN) designed for this task, has demonstrated the ability to identify phenotype-associated cell populations at frequencies as low as 0.01\%. Training such models reliably requires patient cohorts that are larger and more diverse than any single institution can typically assemble, and the underlying single-cell data is too sensitive to share across institutional boundaries under existing privacy regulations. 
We propose a secure multi-party computation (MPC) framework that enables the training and inference of CellCnn entirely on secret-shared data. This ensures that neither the participants nor the computing servers ever observe raw patient data or intermediate values. Evaluated on benchmark single-cell datasets for cytomegalovirus infection (CMV) and acute myeloid leukaemia (AML), our implementation preserves accuracy close to its plaintext counterpart while outperforming the prior privacy-preserving baseline. In contrast to earlier privacy-preserving approaches that removed components such as ReLU activations and bias terms, our method retains these key parts of the CellCnn architecture and supports accurate analysis without exposing raw patient data.}

\begin{document}

\renewcommand{\thefootnote}{}
\footnotetext{\small{Article version: \datedate $\;$ h\currenttime  $\;$ CET}}

\thispagestyle{myheadings}
\pagestyle{myheadings}
\markright{\tt Proceedings of CIBB 2026}

\section{Introduction}
\label{sec:SCIENTIFIC-BACKGROUND}

Rare cell populations play a central role in the initiation and progression of many diseases. Tumour-initiating cells, leukaemic blasts in minimal residual disease (MRD), and memory-like natural killer cells associated with cytomegalovirus (CMV) infection are examples of cell subsets whose detection is clinically important yet technically challenging due to their low frequency~\cite{cellcnn}. Advances in single-cell measurement technologies, particularly mass cytometry, now enable the simultaneous quantification of dozens of protein markers across thousands of individual cells, providing the resolution needed to characterize such rare populations~\cite{Bendall2011MassCytometry}.

CellCnn~\cite{cellcnn} is a CNN specifically designed to detect disease-associated cell subsets from these high-dimensional single-cell measurements. Unlike traditional approaches that separate unsupervised feature extraction from supervised classification, CellCnn jointly learns cell population representations and their association with disease phenotypes. The network takes multi-cell inputs (i.e. groups of single-cell measurements from a patient sample) and learns convolutional filters whose weights correspond to molecular profiles of relevant cell subsets. Through pooling, these filters capture the presence or frequency of disease-associated populations, enabling detection even at extremely low frequencies. CellCnn has been successfully applied to identify paracrine signalling responses, HIV progression-associated populations, rare CMV-associated NK-cell subsets, and leukaemic blasts in MRD-like settings with frequencies down to 0.01\%~\cite{cellcnn}.

A fundamental limitation of single-cell analysis is the requirement for sufficiently large and diverse patient cohorts to train robust models. Rare diseases, by definition, have small patient populations in any individual institution. Multi-centre collaboration is therefore essential to achieve the statistical power needed for reliable classification~\cite{pricell}. However, single-cell data is highly sensitive since it can reveal disease status, immune function, and potentially genetic predisposition. Privacy regulations, including the European General Data Protection Regulation (GDPR)~\cite{gdpr} and the US Health Insurance Portability and Accountability Act (HIPAA)~\cite{hipaa1996}, impose strict constraints on sharing such data across institutional boundaries, and have motivated a recent push toward federated and privacy-preserving learning in digital health~\cite{rieke2020}.

The most related prior work is PriCell~\cite{pricell}, which implements CellCnn under multiparty homomorphic encryption in a federated-learning setting: several data providers hold local datasets and jointly evaluate the network on collectively encrypted weights, with passive security against up to $N{-}1$ colluding parties. To stay within the polynomial-circuit budget imposed by CKKS bootstrapping, PriCell omits the ReLU activation between convolution and pooling. It also removes the bias terms and trains a single sigmoid head against mean-squared error. As a consequence it supports only classification and no regression. Also the missing components reduce the expressivity of the network on harder benchmarks. Its synchronous protocol additionally requires all data providers to remain available throughout training and to exchange ciphertexts.

We propose an alternative based on secure multi-party computation with additive secret sharing over $\mathbb{Z}_{2^{64}}$, in which two computing proxies and a single helper party jointly evaluate the CellCnn architecture on secret-shared data without any party observing plaintext values. The MPC setting allows us to retain ReLU activations and bias terms, and to support continuous outcomes through a $\tanh$-based regression head, none of which are available in PriCell. We evaluate on the CMV (NK-cell) and AML benchmarks used by the CellCnn~\cite{cellcnn} and PriCell~\cite{pricell} studies, and show that the same architecture, when trained under MPC, closely approaches the accuracy of its plaintext counterpart and improves on the prior privacy-preserving baseline.

\section{Methods}\label{sec:METHODS}

\subsection{CellCNN Architecture} \label{sec:CellCNN}
The CellCnn architecture~\cite{cellcnn} consists of three layers: a 1D convolutional layer (CL) with ReLU activation, a pooling layer, and a fully connected (FCL) output layer. Each input sample is a multi-cell input of dimension $n \times m$, where $n$ is the number of cells drawn from a patient sample and $m$ is the number of measured markers. The CL applies h filters of size $1 \times m$, each computing a weighted sum over the marker profile of a single cell. The pooling layer aggregates the filter responses across all $n$ cells in a multi-cell input, and the FCL maps the pooled representation to the output classes. 

\subsection{MPC Framework} \label{sec:MPC}
We adopt a three-party MPC framework~\cite{cecilia} based on 2-out-of-2 additive secret sharing over the ring $\mathbb{Z}_{2^{64}}$. The framework involves three parties: two computing proxies, $\party_0$ and $\party_1$, and a helper party, $\party_2$ or $\helper$. Data owners in our setting are healthcare institutions that hold single-cell datasets. They split each data value into two random shares, one sent to each proxy. Each share individually appears uniformly random and reveals nothing about the original value. The proxies collaboratively compute on the shares: additions are performed locally without communication, while multiplications require interaction assisted by $\party_2$, who supplies precomputed correlated randomness (Beaver triples~\cite{beaver1991}) and never observes unmasked shares of either inputs or intermediates. Real-valued computations, such as those required for neural network weights and cell marker measurements, are supported through fixed-point representation within the ring.
The framework operates under a semi-honest security model with an honest majority, where at most one party may be corrupted and parties do not collude. Under this assumption, raw patient data, model updates, and intermediate activations remain secret-shared throughout the computation. Only the agreed output is reconstructed at the end. In our experiments, this is the final prediction or evaluation result. Leakage from the released output itself, such as membership inference or model inversion, is outside the MPC threat model.

\subsection{Adaptations for MPC} \label{sec:MPC-Adaptations}

Adapting CellCnn for a secure context required replacing expensive non-polynomial operations with efficient secure equivalents that are better suited for privacy-preserving computation.

\textbf{Optimiser.} We replace CellCnn's Adam optimiser, which requires per-parameter division and square-root operations, with mini-batch SGD with momentum: an update consisting only of local operations. Bias terms are retained in both convolutional and fully connected (FCL) layers.

\textbf{Hidden activation and pooling.} The CL is followed by a ReLU, which the framework supports natively. We use mean pooling, which needs no communication between parties. Max pooling is also supported, but gave comparable accuracy at the added cost of secure comparisons.

\textbf{Output activation.} The softmax/cross-entropy head is replaced by a per-class sigmoid trained one-vs-rest, approximated by the same degree-three least-squares polynomial $\sigma(z) \approx 0.5 + 0.23323\,z - 0.00981\,z^{3}$ on $z \in [-3, 3]$ used by PriCell~\cite{pricell}. The FCL output is clipped to $[-3, 3]$ before the sigmoid to keep the cubic approximation stable under fixed-point arithmetic.

\textbf{Regression head.} For the regression task, the sigmoid is replaced by a $\tanh$ output trained against mean-squared error, evaluated via a masked secure exponential and division. 

\subsection{Data} \label{sec:Data}

We evaluate our approach on three datasets: two classification benchmarks from CellCnn~\cite{cellcnn} and PriCell~\cite{pricell}, and a third selected to test our implementation's regression capability.

\textbf{Cytomegalovirus infection (CMV)}: A mass cytometry dataset from Strauss-Albee~et~al.~\cite{straussalbee2015} comprising samples from 20 donors with 37 markers per cell, with 11 CMV$^-$ and 9 CMV$^+$ labels. We use 14 donors for training and 6 for testing, following the protocol of~\cite{cellcnn}.

\textbf{Acute myeloid leukaemia (AML)}: A mass cytometry dataset with 16 markers per cell, used for three-class classification (healthy, cytogenetically normal, and core-binding factor translocation). The training and test sets comprise 7 samples and 6 samples respectively following~\cite{cellcnn}.

\textbf{AML minimal-residual-disease (MRD) regression}: We extend the AML benchmark to a continuous-target regression by spiking CBF leukaemic blasts from Levine~et~al.~\cite{levine2015} into healthy bone-marrow samples at log-uniformly drawn fractions $f \in [0.1\%, 5\%]$, with $10\%$ of training samples clamped to $f=0$ as pure-healthy anchors. Donor splits are patient-disjoint between training and held-out test pools, and the spike-in construction provides exact ground-truth $f$, mirroring the rare-cell-detection benchmark protocol of~\cite{cellcnn}.

\section{Results} \label{sec:RESULTS}

We compare three implementations: our MPC implementation of CellCnn (\textit{Ours}), the CellCnn baseline~\cite{cellcnn} (\textit{CellCnn}), and the HE-based PriCell~\cite{pricell}. For each dataset we evaluate two complementary metrics from the CellCnn protocol: \textit{multi-cell input} accuracy, computed over many multi-cell inputs drawn from each sample, and \textit{phenotype} accuracy, computed once per sample by aggregating the predictions of its multi-cell inputs. All implementations are trained on identical train/test splits with identical mini-batch order so that any difference in accuracy reflects only the network or the cryptographic backend. We report mean $\pm$ standard deviation over the ten splits (the standard error of each mean is smaller by $\sqrt{10}$), retaining the std because the split-to-split spread is itself informative for these small cohorts.

\subsection{Cytomegalovirus (CMV/NK)}
\label{sec:results_nk}

\begin{figure}[t]
    \centering    \includegraphics[width=0.65\linewidth]{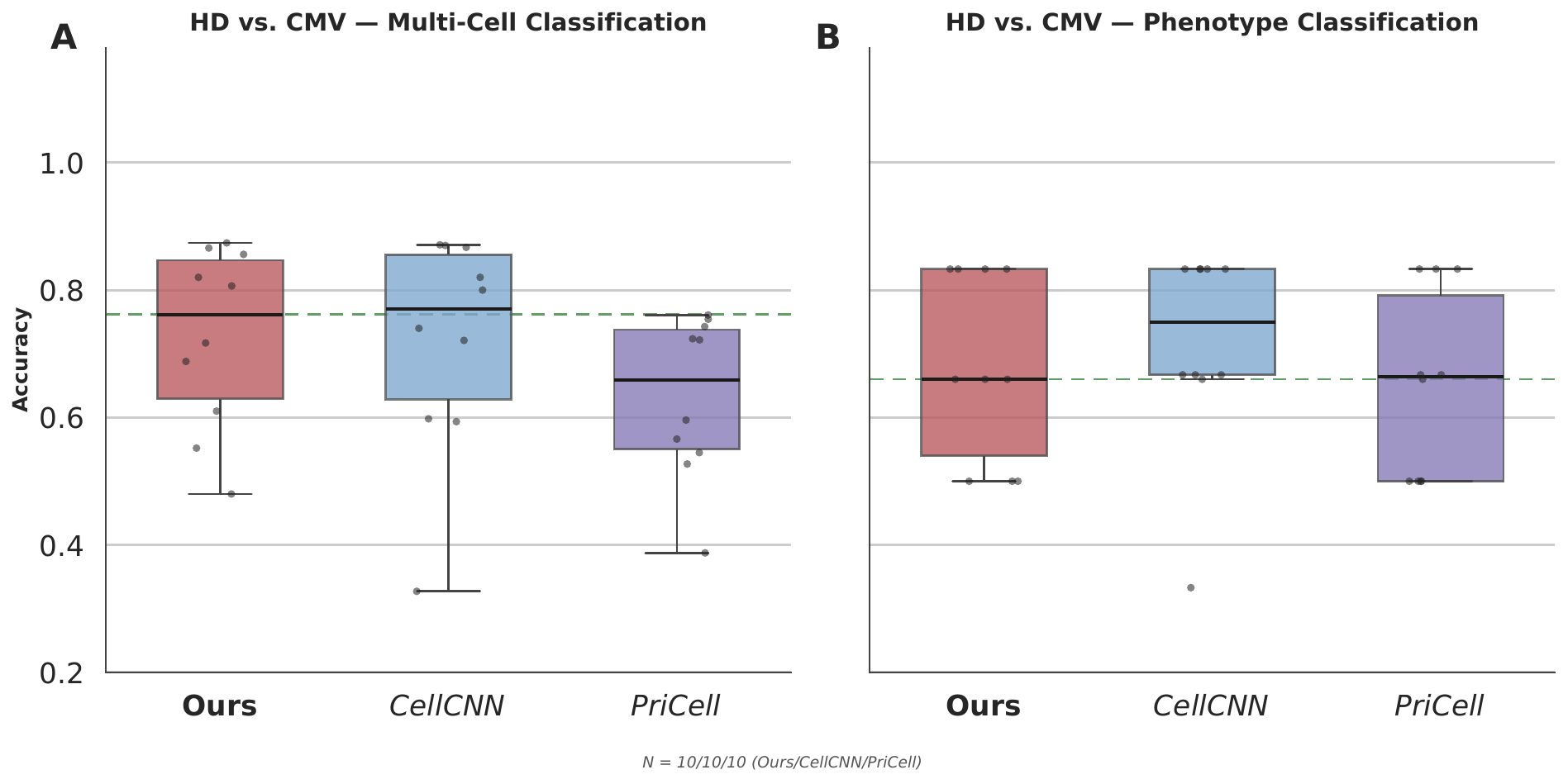}
    \caption{Classification accuracy on the CMV/NK dataset across 10 independent train/test splits. Left: multi-cell-input accuracy. Right: phenotype accuracy. Our MPC implementation achieves accuracy comparable to plaintext CellCnn, while PriCell trails both, consistent with the cost of its polynomial-only architecture.}
    \label{fig:result_nk}
\end{figure}

We run each method on 10 independent train/test splits and report mean and standard deviation in Figure~\ref{fig:result_nk}. Our MPC implementation closely tracks the plaintext CellCnn baseline on both metrics: multi-cell-input accuracy of $0.727 \pm 0.141$ versus $0.721 \pm 0.173$ for plaintext CellCnn, and phenotype accuracy of $0.681 \pm 0.146$ versus $0.716 \pm 0.158$. The differences between the two implementations are well within one standard deviation across splits, indicating that the MPC adaptations of Section~\ref{sec:MPC-Adaptations} do not measurably degrade classification accuracy. PriCell trails both at $0.633 \pm 0.127$ multi-cell-input accuracy and $0.649 \pm 0.146$ phenotype accuracy, a gap consistent with the more restricted architecture that its CKKS bootstrapping budget enforces (no bias, no hidden-layer ReLU, single sigmoid head). A Wilcoxon signed-rank test across the ten paired splits confirms that the multi-cell-input improvement over PriCell is significant ($p{=}0.004$, nine of ten splits favour our method). At the phenotype level the two methods are statistically indistinguishable, as the metric is quantised over six test donors and most splits tie. The large standard deviations ($\pm 0.13$-$0.17$) stem from the small cohort (six donors held out per split, so one misclassified donor shifts accuracy by ${\approx}0.17$) and affect all three methods similarly, so they should be read as split-averaged trends rather than precise point estimates.

\subsection{Acute myeloid leukaemia (AML)}
\label{sec:results_aml}

\begin{figure}
    \centering
    \includegraphics[width=0.65\linewidth]{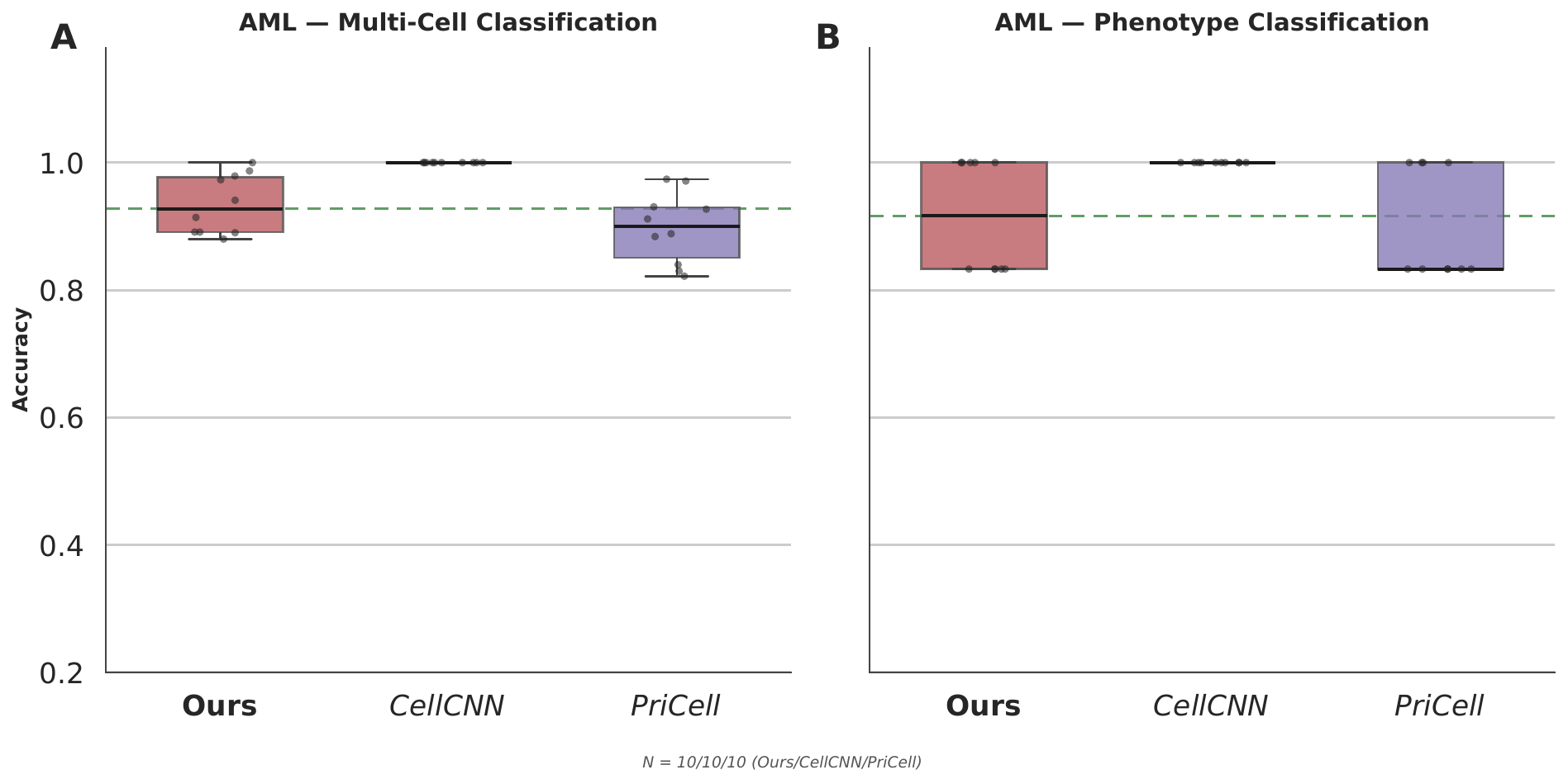}
    \caption{AML 3-class classification accuracy across ten independent splits. Plaintext CellCnn reaches ceiling performance, our MPC implementation retains high accuracy and remains closer to plaintext CellCnn than PriCell.}
    \label{fig:result_aml}
\end{figure}

The AML benchmark is a three-class classification task with only seven training samples, and is known to be highly separable~\cite{cellcnn}. We evaluate each method across ten independent train/test splits (Figure~\ref{fig:result_aml}). CellCnn achieves a perfect $1.000 \pm 0.000$ multi-cell-input accuracy on every split. Our MPC implementation tracks this ceiling closely at $0.935 \pm 0.047$ multi-cell-input accuracy and $0.916 \pm 0.088$ phenotype accuracy. PriCell trails our implementation by roughly three percentage points at $0.898 \pm 0.055$ multi-cell-input and $0.900 \pm 0.086$ phenotype accuracy. The multi-cell advantage over PriCell is consistent ($8/10$ splits) though only marginal at ten splits (Wilcoxon $p{=}0.08$).The gap is consistent across splits and again reflects the architectural restrictions of the homomorphic encryption setting.

\subsection{AML MRD-load regression}
\label{sec:results_regression}

On a patient-held-out test set of $42$ multi-cell inputs (seven spike-in frequencies in $[0\%, 5\%]$, six replicates each), our MPC implementation recovers the underlying CBF blast fraction with Pearson $r = 0.98$ and mean absolute error (MAE) of $1.0$ percentage points (pp). The CellCnn baseline achieves $r = 0.99$ and MAE of $0.8$ pp. Secure execution introduces only a small loss relative to the plaintext baseline.

\subsection{Runtime}
\label{sec:results_runtime}

Our MPC implementation trains the CMV/NK model for 20 epochs in approximately $1{,}200$~s on a 5~Gbps and 0.8ms~RTT LAN between three dedicated MPC parties. The number of computing parties is fixed and independent of the data-owning institutions, who only provide secret shares and need not remain online during training. Throttling to 1~Gbps (RTT unchanged) increases training time to approximately $2{,}700$~s, indicating that bandwidth is a major bottleneck. For reference, PriCell~\cite{pricell} reports approximately $1{,}550$~s for 20 epochs on the same benchmark with $N=10$ active parties on a 1~Gbps LAN; the comparison should be read against different deployment models, as PriCell distributes computation across active data holders while ours uses a fixed set of dedicated parties and tolerates unbalanced contributions.

\section{Conclusion}
\label{sec:CONCLUSIONS}

We have presented an end-to-end implementation of CellCnn under secure MPC that performs both training and inference on additively secret-shared data, exposing no party to plaintext records or intermediate values. By replacing only the strictly non-polynomial primitives of the original network, such as the Adam optimizer, the softmax/cross-entropy head, with their secure-arithmetic counterparts, our implementation retains the ReLU layer, and the multi-class head that prior privacy-preserving CellCnn implementations had to omit. On the CMV/NK and AML benchmarks our MPC implementation preserves high accuracy relative to the plaintext baseline across independent train/test splits. On the AML MRD-load regression task it additionally extends the privacy-preserving CellCnn pipeline to a continuous prediction setting that homomorphic-encryption-based predecessors did not support, recovering the unconstrained baseline's ranking quality on patient-held-out test donors. Several limitations remain: the semi-honest, non-colluding honest-majority model could be strengthened to malicious security. Leakage from the released output lies outside MPC and motivates pairing the pipeline with differential privacy. Also, the benchmarks inherit small cohorts and a synthetic regression ground truth, so validation on larger real cohorts and continuous clinical biomarkers is a natural next step.

\section*{Conflict of interests}
\label{sec:CONFLICT-OF-INTERESTS}
The authors declare no competing interests.

\section*{Acknowledgments}
\label{sec:ACKNOWLEDGMENTS}
The authors thank Prof. Manfred Claassen and the German Network for Bioinformatics Infrastructure (de.NBI) for their support.

\section*{Funding}
\label{sec:FUNDING}
This research is funded by the German Federal Ministry of Education and Research (BMBF) under project number 01ZZ2010 (MDPPML). 

\section*{Availability of data and software code}
\label{sec:AVAILABILITY}
Our software code is available at 
\url{https://github.com/mdppml/CECILIA-CellCNN.git}. 
The CMV and AML datasets are publicly available from the original CellCnn study~\cite{cellcnn}.

\footnotesize
\bibliographystyle{unsrt}
\bibliography{bibliography_CIBB_file.bib} 
\normalsize

\end{document}